\documentclass{ifacconf}
\usepackage[cmex10]{amsmath}

\usepackage{graphicx}      
\usepackage{natbib}        
\begin{document}
\begin{frontmatter}

\title{ PID2018 Benchmark Challenge: learning feedforward control
\thanksref{footnoteinfo}}

\thanks[footnoteinfo]{Corresponding author: Professor YangQuan Chen ({\tt yqchen@ieee.org}). Y. Zhao and J. Yuan are supported by China Scholarship Council.}

\author[Third,First]{Yang Zhao},
\author[First]{Sina Dehghan},
\author[First,Fifth]{Abdullah Ates},
\author[First,Fourth]{Jie Yuan},
\author[Third]{Fengyu Zhou},
\author[Third]{Yan Li}, and
\author[First]{YangQuan Chen}

\address[Third]{School of Control Science and Engineering, and Center for Robotics, Shandong University, Jinan 250061, Shandong, P. R. China (e-mail:{ \tt zdh1136@gmail.com}).}

\address[First]{University of California Merced, Merced, CA 95340 USA (e-mail: { \tt \{sdehghan,ychen53\}@ucmerced.edu}).}

\address[Fourth]{School of Automation, Southeast University, Nanjing
210096, China (e-mail: {\tt{jieyuan@seu.edu.cn}}).}
\address[Fifth]{Engineering Faculty, Computer Engineering Department, Inonu University,  44280, Malatya, Turkey (e-mail: { \tt abdullah.ates@inonu.edu.tr}).}

\begin{abstract}                
The design and application of learning feedforward controllers (LFFC) for the one-staged refrigeration cycle model described in the PID2018 Benchmark Challenge is presented, and its effectiveness is evaluated. The control system consists of two components: 1) a preset PID component and 2) a learning feedforward component which is a function approximator that is adapted on the basis of the feedback signal. A B-spline network  based LFFC and a low-pass filter based LFFC are designed to track the desired outlet temperature of evaporator secondary flux and the superheating degree of refrigerant at evaporator outlet. Encouraging simulation results are included. Qualitative and quantitative comparison results evaluations show that, with little effort, a high-performance control system can be obtained with this approach.
Our initial simple attempt of low-pass filter based LFFC and B-spline network based LFFC give J=0.4902 and J=0.6536 relative to the decentralized PID controller, respectively. Besides, the initial attempt of a combination controller of our optimized PI controller and low-pass filter LFFC gives J=0.6947 relative to the multi-variable PID controller.
\end{abstract}


\end{frontmatter}

\section{Introduction}
Dates backs to centuries, the research of refrigeration systems has experienced worldwide significant advances introduced by industry and research institutes. The purpose of refrigeration is to attain and maintain a temperature below that of the surroundings, the aim being to cool some product or space to the required temperature. Working in the same way, air conditioning and refrigeration systems are extensively applied in food preservation, chemical and process industries, manufacturing processes, cold treatment of metals, drug manufacture, ice manufacture and above all in areas of industrial air conditioning and comfort air conditioning. After more than 100 years of design evolution, vapor compression refrigeration systems are now the most common means for commercial and residential space cooling, which brings a problem of large deal of energy consumption and negatively energy and economic balances affects (\cite{main}).

In recently years, a large body of work has been generated to adapt linear techniques for control which can be found in the literature are decentralized PID control (\cite{underwood2001analysing}; \cite{wang2007study}; \cite{marcinichen2008dual}; \cite{salazar2014pid}), decoupling multivariable control (\cite{shen2010normalized}), LQG control (\cite{he1996dynamic}; \cite{schurt2009model}; \cite{schurt2010assessment}), model predictive control (MPC) (\cite{razi2006neuro}; \cite{sarabia2009hybrid}; \cite{ricker2010predictive}; \cite{fallahsohi2010predictive}), and robust $H_\infty$ control (\cite{larsen2003modelling}; \cite{bejarano2015multivariable}). Whereas, there are many challenges associated with refrigeration systems control stemming from the components themselves to the fundamental characteristics of a heat transfer process, which cause high thermal inertia, dead times, high coupling between variables, and strong nonlinearities. Therefore, a less accurate model of the plant will result in a controller with an unsatisfied performance. When the model is not available or when many parameters cannot be determined, learning feedforward control (LFFC) may be considered.

Being a class of iterative learning control (ILC) (\cite{6arimoto1984bettering}; \cite{7moore2012iterative}), learning feedforward control (LFFC) shares basic ideas with ILC (\cite{1starrenburg1996learning}; \cite{2velthuis1996learning}; \cite{3otten1997linear}; \cite{4velthuis2000stability}; \cite{5velthuis2000learning}). First proposed for motion systems subjected to reproducible disturbances, LFFC is designed to compensate the reproducible disturbances as value-added blocks (\cite{velthuis2000regularisation}).
As an extra degree of freedom, LFFC generates steering signals that enhances the feedback control performance (\cite{Main_chen2004learning}). Many previous research (\cite{boeren2015optimal}; \citep{taherkhani2011lffc}; \citep{lin2011learning}) had shown that learning feedforward control can improve system performance and acquire enhanced extrapolation capabilities for repetitive tracking control tasks with little modeling information. The main contribution of this paper is to apply two-parameter tunable LFFC schemes for the control of refrigeration systems introduced in PID2018 benchmark problem (\cite{main}).

We first discuss the one-stage vapour-compression refrigeration system in more detail in Section 2. Next, the design of the learning feedforward control system is discussed (Section 3). Simulation results are presented in Section 4. We end with conclusions in Section 5.

\section{PID2018 Benchmark Challenge Introduction}
\begin{figure}[!htb]
  \centering
  \includegraphics[width=0.8\hsize]{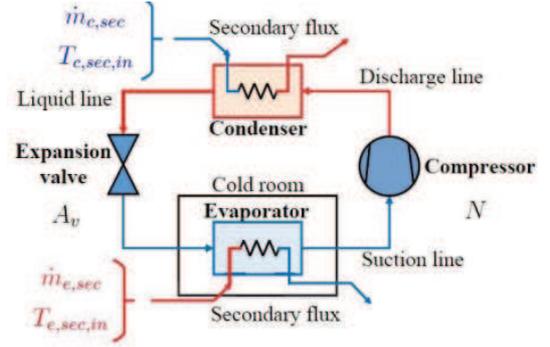}
  \caption{The refrigeration cycle}
  \label{fig:refrigertaion_cycle}
\end{figure}

\begin{figure}
\begin{center}
\includegraphics[width=8.4cm]{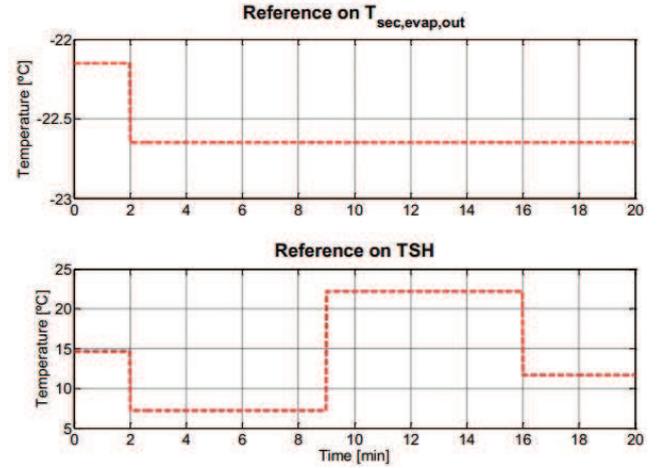}
\caption{The standard simulation for Benchmark PID 2018 generates changes in the references $T_{e,sec,out}$ and $T_{SH}$.}
\label{fig:reference_desired}
\end{center}
\end{figure}

\begin{figure}
\begin{center}
\includegraphics[width=8.4cm]{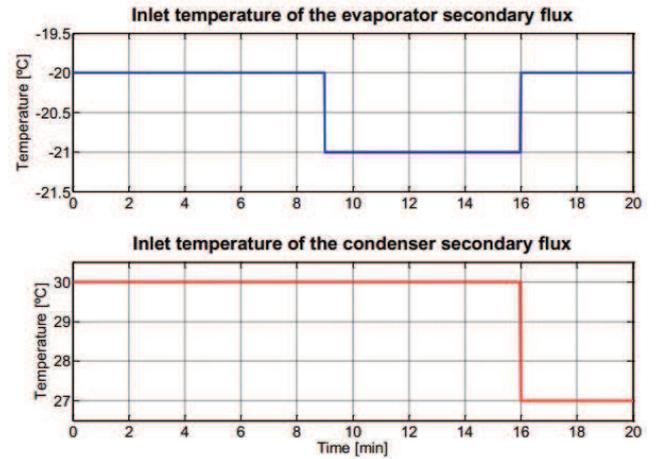}
\caption{The standard simulation for Benchmark PID 2018 generates changes in two disturbances: $T_{e,sec,in}$ and $T_{c,sec,in}$.}
\label{fig:disturbance_plot}
\end{center}
\end{figure}

In this section we describe the control problem introduced in PID2018 Benchmark Challenge (\cite{main}) in brief to introduce the necessary information for the control system design process. A one-stage vapor-compression refrigeration cycle model (Fig. \ref{fig:refrigertaion_cycle}) is to be controlled for this challenge. The model is consisted of evaporator, condenser, variable speed compressor, and expansion valve where the expansions valve's opening $A_v$ (percentage) and compressor speed $N$ (Hz) are to be manipulated to control the outlet temperature of evaporator secondary flux, $T{sec\_evap\_out}$ ($ ^{\circ} C$), and the degree of superheating of the refrigerant at the evaporator outlet, $TSH$ ($ ^{\circ} C$). Hence, our control objective is to ensure these two variables to track their references as efficiently as possible in presence of disturbances by manipulating the compressor speed and the expansion valve opening. Therefore the whole control system would be a two-input, two-output system. Fig. \ref{fig:reference_desired} shows the desired reference for the controlled variables of our challenge. Besides, the inlet temperature of the evaporator secondary flux and the inlet temperature of the condenser secondary flux are changed as shown in Fig. \ref{fig:disturbance_plot} to add disturbance to the system.
It is important to note that the manipulated variables, $A_v$ and $N$, are subjected to limits, $A_v \in [10, 100]$ and $N \in [30, 50]$, and are saturated within the system block.

\section{Design and Implementation of Learning Feedforward Component}

\begin{figure}
\begin{center}
\includegraphics[width=0.8\hsize]{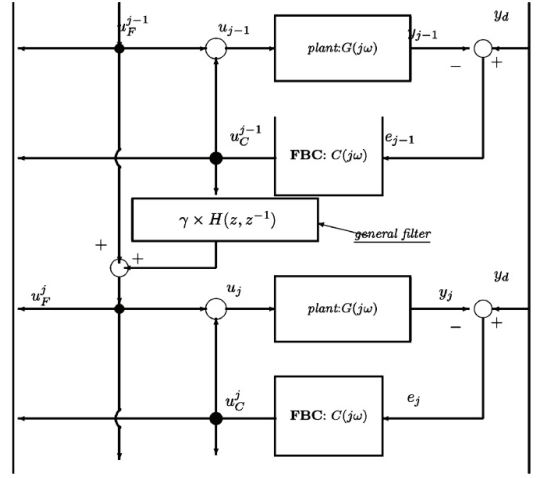}    
\caption{Block diagram of LFFC with a general filter $H(z,z^{-1})$(\cite{Main_chen2004learning}).}
\label{fig:Fig1}
\end{center}
\end{figure}

The learning feedforward controller works in the sense that a nonlinear mapping between the reference input(s) and the force output can be created by capitalizing on systems repetitiveness. Fig. \ref{fig:Fig1} shows the general LFFC scheme. Here $y_d$ and FBC denote desired output trajectory and  "feedback controller" respectively. To construct the control signal for the repetition number j, $U_F^j$, the feedforward control signal $U_F^{j-1}$ and the feedback control signal $U_C^{j-1}$ for the (j-1)th operation are stored in the memory bank.
Thus, the overall control signal at the $j$th repetitive operation is adapted according to
$$u^{j}(t)=u^{j}_{C}(t)+u^{j}_{F}(t).$$
As illustrated in Fig. \ref{fig:Fig1}, the $j$th iteration feedforward control is generated with the combination of last feedforward control and filtered evaluations of feedback control.
\begin{equation}\label{uf_equation}
u^{j}_{F}=u^{j-1}_{F}+\gamma H(z,z^{-1})u_{C}^{j-1}.
\end{equation}
where $H(z,z^{-1})$ is the filter, $\gamma$ represents the learning gain, $j$ denotes the $j$th repetitive operation.
Proper selection of filter $H(z,z^{-1})$ and the learning parameter $\gamma$ plays the key role in design of LFFC. A number of design choices of the two undetermined parts has been tried which are illustrated as following.

\subsection{B-spline network (BSN)}

\begin{figure}
\begin{center}
\includegraphics[width=0.8\hsize]{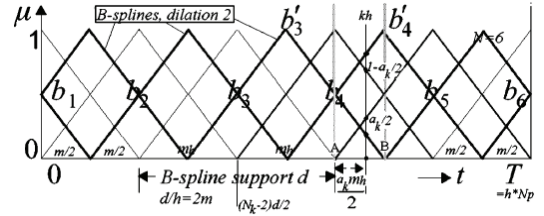}
\caption{Second-order dilated B-splines and the filtering process(\cite{Main_chen2004learning}).}
\label{fig:second_order_BSN}
\end{center}
\end{figure}

The approach we first considered in this paper as the feedforward component is a B-splines network (BSN). In general, a BSN is a function
$$y=f(x)=\sum_{i}\omega_{i}\mu_{i}(x)$$
where $y$, $x$, $\mu_{i}(\cdot)$ and $\omega_{i}$ are the BSN output and input, membership function, and B-spline weights, respectively. Comparable to radial basis function networks, BSNs are one-hidden-layer networks with updated weights between the hidden layer and output layer. The region of the input space on which the basis functions are nonzero is called the support of the basis, is $d$. Generally, the support of a membership function neither equal to the whole input space nor overlap half. Besides, dilated BSN is used to define the case that the supports of the basis functions do overlap more than half of the input space. As illustrated in Fig. \ref{fig:second_order_BSN},  there are $2m=d/h$ samples within one B-spline if the sampling period is $h$. To create the feedforward control, the inputs to the neural network generally include the feedback control signal $u_{C}(t)$, time $t$ or state $x(t)$, and reference or desired trajectory $y_{d}(t)$. For our purpose, we choose the time t as the input space. In this paper, following the lead from paper \cite{Main_chen2004learning}, we consider a second-order dilated BSN, illustrated in Fig. \ref{fig:second_order_BSN}, whose basis functions are piecewise polynomial functions of order $1$. The overall learning filter $H(z,z^{-1})$ in equation (\ref{uf_equation}) is that
$$H(z,z^{-1})=A_{2}(\omega,a_{k},d)\left(\frac{sin(\omega d/4)}{\omega d/4}\right)^{2}$$
Please refer to (\cite{Main_chen2004learning}) for more details about the stability analysis of this method and design formula for the only two coefficients, the B-spline support width $d$ and the learning gain $\gamma$.


\subsection{Low-pass filter (LPF)}
Our goal is to add learning feedforward control to the system. A standard simulation, starting from prepared operating point described, has been scheduled for the Benchmark PID 2018. As shown in Fig. \ref{fig:disturbance_plot}, the simulation includes step changes in the most important disturbances: the inlet temperature of the evaporator secondary flux $T_{e,sec,in}$, and the inlet temperature of the condenser secondary flux $T_{c,sec,in}$. It is notable that the disturbances is relatively clean and simple. In such case, BSN based LFFC may exerts not very well, thus another LFFC based on low-pass filter is proposed to achieve better tracking performance. This learning feedforward control output may be written as
\begin{equation}\label{LPF_equation}
u^{j}_{F}=u^{j-1}_{F}+\gamma e^{-\frac{t}{\tau}}u_{C}^{j-1}.
\end{equation}
The stored feedback control signal $u_{C}^{j-1}$ is filtered through a low-pass filter and multiplied by a learning gain $\gamma$, where there are also only two parameters to be tuned.

\textit{Guidelines for Tuning} The low-pass filter based LFFC scheme offers considerable flexibility with two tunning knobs that it provides. In the current scheme the learning gain $r$  and filter parameter $\tau$ in the learning function, equation (\ref{LPF_equation}), were tuned  by try and error in the direction that reduces the overall index mentioned in section 4.

\section{Results and discussions}
In this section, simulation studies of our proposed learning feedforward schemes are demonstrated. The qualitative and quantitative comparisons are explored between our designed control scheme with discrete decentralized PID (labelled as Controller 1) and multi-variable PID (labelled as Controller 2) provided in the Benchmark PID 2018. Simulations were performed with the MATLAB program RS\_simulation\_management.m to demonstrate feasibility of learning control. The sampling time is 1s and the simulation time is 1200s. Moreover, eight individual performance indices and one combined index are applied to further evaluate in comparison.

\begin{figure}[!htb]
  \centering
  \includegraphics[width=\hsize]{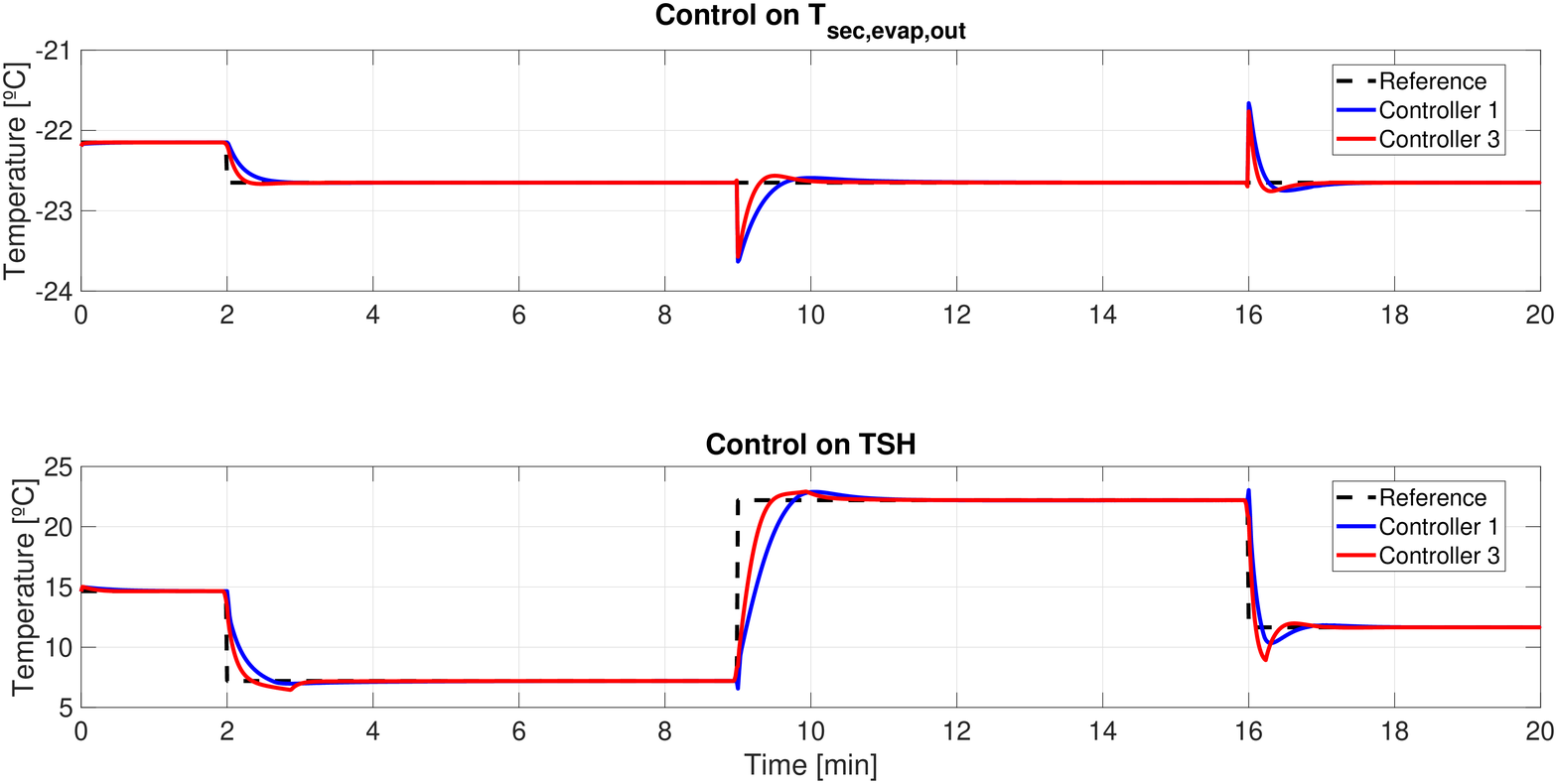}
  \caption{Tracking performance at 10th iteration compared with Controller 1 under BSN LFFC.}
  \label{fig:BSN_M9_Comparison_Control}
\end{figure}

\begin{figure}[!htb]
  \centering
  \includegraphics[width=0.7\hsize]{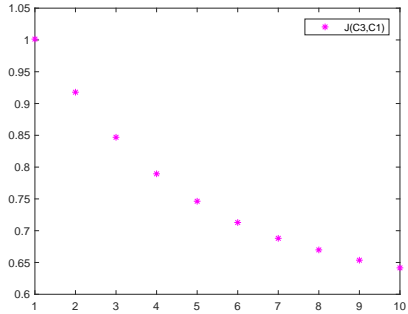}
  \caption{The change of combined index with iterations under BSN LFFC.}
  \label{fig:BSN_M9_J}
\end{figure}

\begin{figure}[!htb]
  \centering
  \includegraphics[width=\hsize]{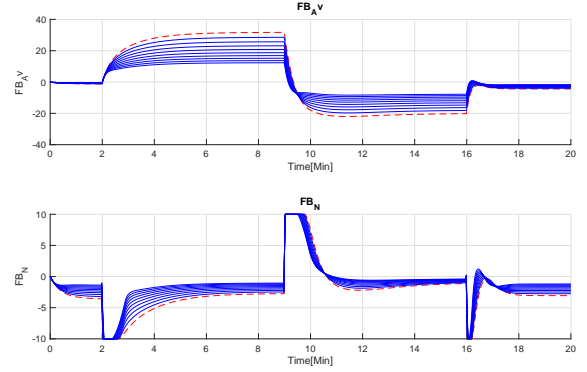}
  \caption{The change of feedback control with iterations under BSN LFFC.}
  \label{fig:BSN_M9_FeedBackchange}
\end{figure}

For a dilated BSN LFFC with dilation 2, m=9 and $\gamma$=0.1, labelled as Controller 3, Fig. \ref{fig:BSN_M9_Comparison_Control}-Fig. \ref{fig:BSN_M9_FeedBackchange} show its tracking performance after 10th learning iterations compared with the discrete decentralized PID controller (labelled Conttroller 1). As shown in Fig. \ref{fig:BSN_M9_Comparison_Control}, Controller 2 achieves tighter control on the outlet temperature of the evaporator secondary flux and the degree of superheating than Controller 1. Fig. \ref{fig:BSN_M9_J} depicts the convergence of our algorithm with combined index $J$. In Fig. \ref{fig:BSN_M9_FeedBackchange}, the red line depicts the feedback control signal with the reference controller, and the blue lines represent the feedback control signal with our BSN learning feedforward controller. As the blue lines go smaller, the feedforward controller undertakes more work with iterations. The eight performance indices shown in Table \ref{tb:indices1} further testify the control effort in BSN LFFC.
\begin{table}[hb]
\begin{center}
\caption{Quantitative Comparison of BSN LFFC with decentralized controller}\label{tb:indices1}
\begin{tabular}{ccc}
index & C3 vs C1  \\\hline
$RIAE_1$  & 0.5389   \\
$RIAE_2$  & 0.6068   \\
$RITAE_1$  & 0.6915   \\
$RITAE_2$  & 0.9157   \\
$RITAE_2$  & 0.5753   \\
$RITAE_2$  & 0.6583   \\
$RIAVU_1$  & 1.0383   \\
$RIAVU_2$  & 1.0514   \\\hline
$J$ & 0.6536   \\ \hline
\end{tabular}
\end{center}
\end{table}

\begin{figure}[!htb]
  \centering
  \includegraphics[width=\hsize]{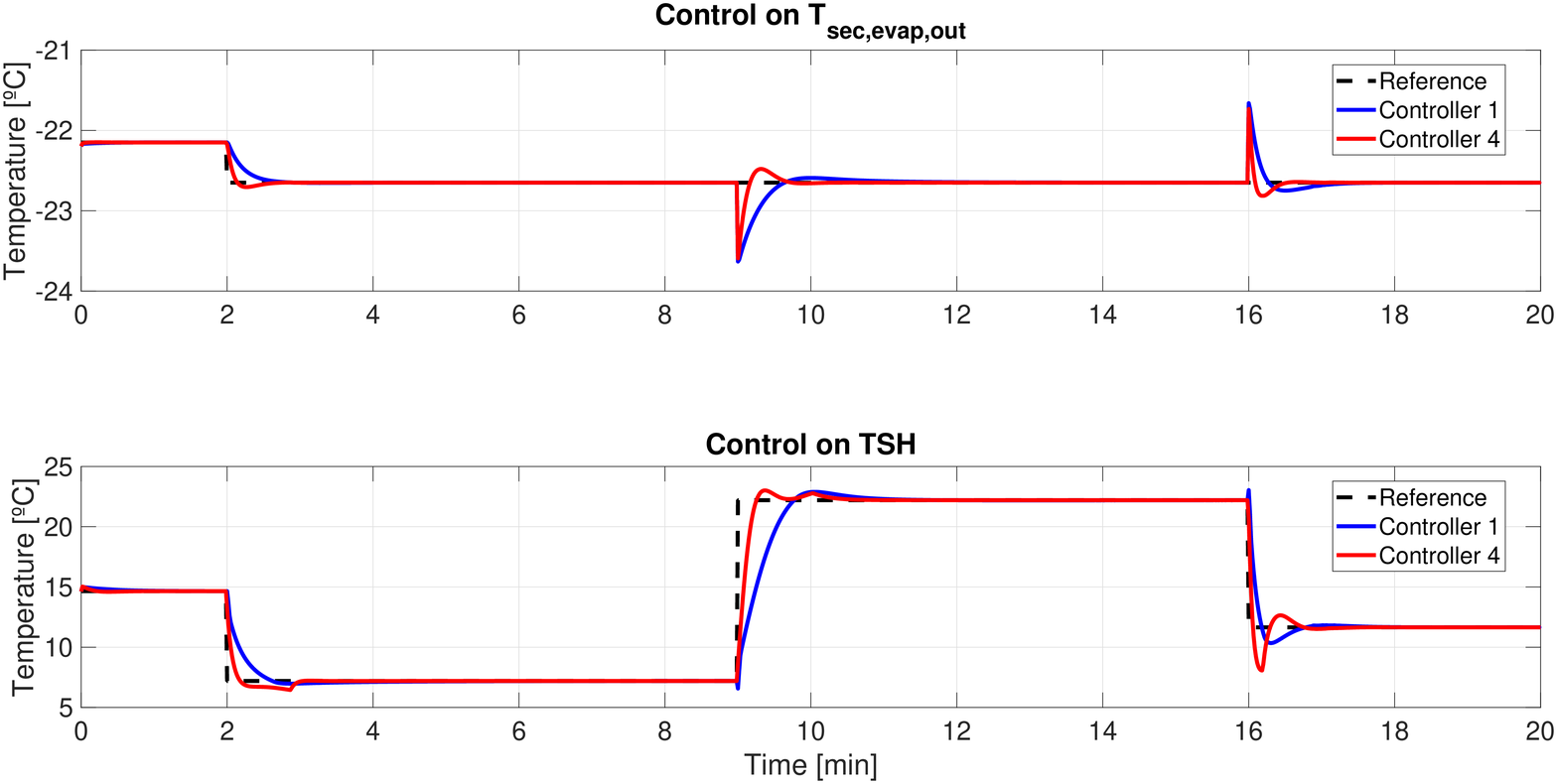}
  \caption{Tracking performance at 20th iteration compared with Controller 1 under LPF LFFC.}
  \label{fig:Comparison_Control}
\end{figure}

\begin{figure}[!htb]
  \centering
  \includegraphics[width=0.7\hsize]{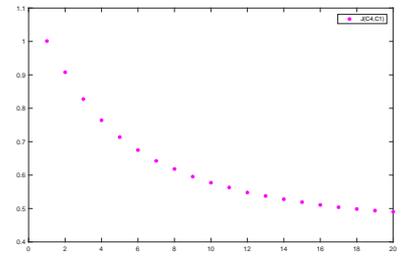}
  \caption{The change of combined index with iterations under LPF LFFC.}
  \label{fig:J_FB}
\end{figure}

For low-pass filter based LFFC,  a simple limit version (with $\gamma$=0.1, $\tau=0$) labelled as Controller 4 is applied. Comparison results with the discrete decentralized PID is shown in Fig. \ref{fig:Comparison_Control}. As shown in Fig. \ref{fig:J_FB}, the change of combined index J illustrates the convergence of this scheme. The performance indices in Table \ref{tb:indices2} show that the proposed low-pass filter learning feedforward controller can achieve better tracking performance.

\begin{table}[hb]
\begin{center}
\caption{Quantitative Comparison of LPF LFFC with decentralized controller}\label{tb:indices2}
\begin{tabular}{ccc}
index & C4 vs C1  \\\hline
$RIAE_1$  & 0.4417   \\
$RIAE_2$  & 0.4880   \\
$RITAE_1$  & 0.3170   \\
$RITAE_2$  & 0.3832   \\
$RITAE_2$  & 0.2558   \\
$RITAE_2$  & 0.7937   \\
$RIAVU_1$  & 1.0782   \\
$RIAVU_2$  & 1.1631   \\\hline
$J$ & 0.4902   \\ \hline
\end{tabular}
\end{center}
\end{table}

\begin{figure}[!htb]
  \centering
  \includegraphics[width=\hsize]{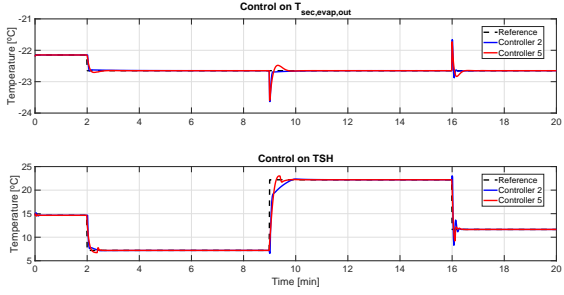}
  \caption{Tracking performance compared with Controller 2 under our combined controller.}
  \label{fig:optimized_Compare}
\end{figure}

\begin{figure}[!htb]
  \centering
  \includegraphics[width=0.7\hsize]{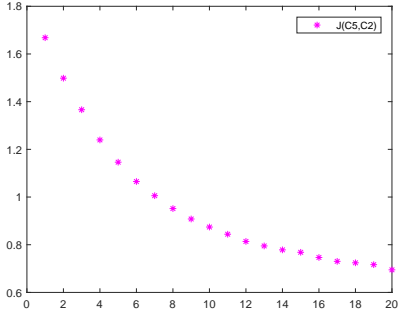}
  \caption{The change of combined index with iterations under our combined controller.}
  \label{fig:optimized_J}
\end{figure}

To further testify the effectiveness of our proposed LFFC, the above mentioned low-pass filter LFFC combined with optimized PI controller proposed in (\cite{Abdullah2018Benchmark}) is applied, which is labelled as Controller 5. Comparison results with the multi-variable PID is shown in Figs. \ref{fig:optimized_Compare}. Fig. \ref{fig:optimized_J} depicts the convergence of this controller. Table 3 shows the performance indexes calculated for all the  PID controller provided in (\cite{main}).

\begin{table}[hb]
\begin{center}
\caption{Quantitative Comparison of C5 with C2 and C1}\label{tb:indices3}
\begin{tabular}{ccc}
index & C5 vs  C2  & C5 vs  C1\\\hline
$RIAE_1$  & 1.2542    & 0.4403\\
$RIAE_2$   & 0.7397   & 0.3297 \\
$RITAE_1$   & 0.2221   & 0.3577\\
$RITAE_2$   & 0.4441   & 0.0812\\
$RITAE_2$    & 0.3472   & 0.1110\\
$RITAE_2$    &0.7822    & 0.1001\\
$RIAVU_1$  & 0.9517   &  1.0738\\
$RIAVU_2$  & 1.1497  &  1.5796\\\hline
$J$ & 0.6947        &  0.3744  \\ \hline
\end{tabular}
\end{center}
\end{table}

\begin{ack}
The first author wishes to thank the members of PTUC SIG (Precision Temperature Uniformity Control Special Interest Group) at UC Merced MESA Lab \footnote{{\tt http://mechatronics.ucmerced.edu/ptuc}}
for fruitful discussions and productive meetings. Project supported by National Natural Science Foundation of China (Grant No. 61375084 and Grant No. 61773242), Key Program of Natural Science Foundation of Shandong Province (Grant No. ZR2015QZ08), Key Program of Scientific and Technological Innovation of Shandong Province (Grand No. 2017CXGC0926), Key Research and Development Program of Shandong Province (Grant No. 2017GGX30133), The National Key Research and Development Program of China (Grant No. 2017YFB1302400).
\end{ack}

\subsection{Conclusion}
In this paper, we demonstrated the effectiveness of a learning feedforward control scheme for vapour-compression refrigeration system. Combined with feedback PID controller, the BSN based and LPF based learning feedforward controllers are applied. The learning controller is able to improve system performance drastically with only two parameters to adjust: the filter coefficient and learning gain. Simulation results show that the proposed controllers can achieve satisfied tracking performance and fast error convergence. It is noteworthy that there still have plenty of room for improvement regarding this method as the parameters are not optimally tuned.

\bibliography{ifacconf}             








\end{document}